\documentclass[sigconf,screen]{acmart} 

\AtBeginDocument{%
  }

\setcopyright{rightsretained}
\acmDOI{10.1145/3663529.3663803}
\acmYear{2024}
\copyrightyear{2024}
\acmSubmissionID{fsecomp24demo-p37-p}
\acmISBN{979-8-4007-0658-5/24/07}
\acmConference[FSE Companion '24]{Companion Proceedings of the 32nd ACM International Conference on the Foundations of Software Engineering}{July 15--19, 2024}{Porto de Galinhas, Brazil}
\acmBooktitle{Companion Proceedings of the 32nd ACM International Conference on the Foundations of Software Engineering (FSE Companion '24), July 15--19, 2024, Porto de Galinhas, Brazil}
\received{2024-01-29}
\received[accepted]{2024-04-15}

\usepackage{amsmath,amsfonts}
\usepackage{algorithmic}
\usepackage{algorithm}
\usepackage{array}
\usepackage[caption=false,font=normalsize,labelfont=sf,textfont=sf]{subfig}
\usepackage{textcomp}
\usepackage{stfloats}
\usepackage{url}
\usepackage{verbatim}
\usepackage{graphicx}
\usepackage{balance}
\usepackage{xspace}
\usepackage{hyperref}
\usepackage{listings}
\usepackage{makecell}
\usepackage{booktabs}
\usepackage{algorithm}
\usepackage{caption}
\usepackage{threeparttable}
\usepackage{longtable}
\usepackage{tabularray}
\usepackage{cleveref}
\hypersetup{hidelinks}
\usepackage{color, colortbl}
\usepackage{balance}
\usepackage[inline]{enumitem}
\usepackage{calligra}
\usepackage{amsmath,amsthm}
\usepackage{tikz}
\usetikzlibrary{shapes.geometric, arrows}
\usetikzlibrary{positioning}
\usepackage{pifont}
\usepackage{tcolorbox}
\usepackage{makecell}
\usepackage{fancybox}
\usepackage{microtype}

\makeatletter

\lstdefinestyle{mystyle}{
    backgroundcolor=\color{backcolour},   
    basicstyle=\footnotesize,
    commentstyle=\color{codegreen},
    keywordstyle=\color{magenta},
    numberstyle=\tiny\color{codegray},
    stringstyle=\color{codepurple},
    basicstyle=\ttfamily\footnotesize,
    breakatwhitespace=true,         
    breaklines=true,                 
    captionpos=t,                    
    keepspaces=true,                 
    numbers=left,                    
    numbersep=5pt,                  
    showspaces=false,                
    showstringspaces=false,
    escapeinside={*}{*},
    showtabs=false,                  
    tabsize=2,
    frame=none,
    postbreak=\mbox{\textcolor{red}{$\hookrightarrow$}\space},
}
\lstdefinestyle{tablecode}{
  basicstyle=\small,
  breaklines=true,
  backgroundcolor=\color{white},
  stringstyle=\color{codepurple},
  numbers=none,
  frame=none,
  basicstyle=\ttfamily\scriptsize,
  xleftmargin=0pt,
  framexleftmargin=0pt,
  framexrightmargin=0pt,
  framexbottommargin=0pt,
  framextopmargin=0pt,
  showspaces=false,  
  showtabs=false,   
  showstringspaces=false, 
  escapeinside={|}{|},
  mathescape=true,
  morekeywords={value},
  deletekeywords=int,
}

\definecolor{tearose}{rgb}{0.96, 0.76, 0.76}
\definecolor{ruddypink}{rgb}{0.88, 0.56, 0.59}
\definecolor{pastelred}{rgb}{1.0, 0.41, 0.38}
\definecolor{javapurple}{rgb}{0.5,0,0.35}
\definecolor{teagreen}{rgb}{0.82, 0.94, 0.75}
\definecolor{codegreen}{rgb}{0.1,0.9,0}
\definecolor{yellow-green}{rgb}{0.6, 0.8, 0.2}
\definecolor{codegray}{rgb}{0.5,0.5,0.5}
\definecolor{codepurple}{rgb}{0.58,0,0.82}
\definecolor{backcolour}{rgb}{0.95,0.95,0.92}
\definecolor{blueannoback}{RGB}{234,242,250}
\definecolor{myblue}{rgb}{0, 0, 0.75}
\definecolor{lightgray}{gray}{0.85}
\definecolor{Blue}{rgb}{0.1, 0.1, 0.2}

\newcommand{\smcode}[1]{\texttt{\fontsize{7.5}{8}\selectfont #1}}

\theoremstyle{definition}

\newcommand{\tool}{\emph{EM-Assist}\xspace}

\newcommand*\circled[1]{\tikz[baseline=(char.base)]{
            \node[shape=circle,draw,inner sep=0.8pt] (char) {#1};}}

\lstdefinestyle{javaStyle}{
    language=Java,
    basicstyle=\footnotesize\ttfamily, 
    keywordstyle=\color{blue},
    commentstyle=\color{green!40!black},
    stringstyle=\color{purple},
    numbers=left,
    numberstyle=\tiny\color{gray},
    stepnumber=1,
    numbersep=5pt,
    backgroundcolor=\color{white},
    showspaces=false,
    showstringspaces=false,
    frame=none,
    rulecolor=\color{black},
    tabsize=4,
    captionpos=b,
    breaklines=true,
    breakatwhitespace=false,
    title=\lstname,
    escapeinside={\%*}{*)},
    morekeywords={*,...}
}

\newboolean{showcomments} 
\setboolean{showcomments}{true}
\ifthenelse{\boolean{showcomments}}
{\newcommand{\nb}[2]{
		\fbox{\bfseries\sffamily\scriptsize#1}
		{\sf\small$\blacktriangleright$\textit{\textcolor{red}{#1}}$\blacktriangleleft$}
	}
}
{\newcommand{\nb}[2]{}}

\ifthenelse{\boolean{showcomments}}
{
}
{}

\ifthenelse{\boolean{showcomments}}
{
}
{}

\newcommand{\EM}{\textit{Extract Method}\xspace}

\newcommand{\extendedCorpusRecallEMAssist}{\ecBestToolRecall}
\newcommand{\extendedCorpusRecallJExtract}{39.4\%\xspace}

\usepackage{microtype}

\newcommand{\surveyQoneApprovalRatio}{94.4\%\xspace}
\newcommand{\usabilityRespondents}{18\xspace}

\newcommand{\extendedCorpusSizeToRunCompetitors}{1,752\xspace}

\newcommand{\extendedCorpusSize}{1,752\xspace}

\newcommand{\llmReliabilityPercentageExtendedCorpus}{76.3\%\xspace}
\newcommand{\llmInvalidSuggestionsPercentageExtendedCorpus}{57.4\%\xspace}
\newcommand{\llmNotUsefulSuggestionsPercentageExtendedCorpus}{18.9\%\xspace}

\newcommand{\averageSuggestionsPerFunctionCommunityCorpora}{27\xspace}

\newcommand{\ecBestToolRecall}{53.4\%\xspace}

\newcommand{\qOneAcceptance}{94.4\%\xspace}
\newcommand{\qTwoAcceptance}{77.7\%\xspace}
\newcommand{\qThreeAcceptance}{94.4\%\xspace}

\begin{document}



\title{\tool: Safe Automated ExtractMethod Refactoring with LLMs}

\author{Dorin Pomian*, Abhiram Bellur*, Malinda Dilhara} \author{Zarina Kurbatova, Egor Bogomolov, Andrey Sokolov, Timofey Bryksin, Danny Dig}
\affiliation{%
  \institution{University of Colorado Boulder, JetBrains Research}
  \country{USA, Cyprus, the Netherlands, Serbia}
}

\email{{dorin.pomian, abhiram.bellur, malinda.malwala, danny.dig}@colorado.edu}

\email{{zarina.kurbatova, egor.bogomolov,andrey.sokolov, timofey.bryksin}@jetbrains.com}

\renewcommand{\shortauthors}{Pomian et al.}

\begin{abstract}
Excessively long methods, loaded with multiple responsibilities, are challenging to understand, debug, reuse, and maintain. The solution lies in the widely recognized \EM refactoring. While the application of this refactoring is supported in modern IDEs, recommending which code fragments to extract has been the topic of many research tools.
However, they often struggle to replicate real-world developer practices, resulting in recommendations that do not align with what a human developer would do in real life. 
To address this issue, we introduce \tool, an IntelliJ IDEA plugin that uses LLMs to generate refactoring suggestions and subsequently validates, enhances, and ranks them. Finally, \tool uses the IntelliJ IDE to apply the user-selected recommendation. 
In our extensive evaluation 
of \extendedCorpusSizeToRunCompetitors real-world refactorings that actually took place in open-source projects, EM-Assist’s recall rate was \extendedCorpusRecallEMAssist among its top-5 recommendations, compared to \extendedCorpusRecallJExtract for the previous best-in-class tool that relies solely on static analysis.  
Moreover, we conducted a usability survey with \usabilityRespondents industrial developers and 
\surveyQoneApprovalRatio gave a positive rating.

The source code, datasets, and distribution of the plugin are available 
on GitHub~\cite{tool:sourceCode}.  It is also available to install on JetBrains Marketplace~\cite{tool:Marketplace}.
A demon video and screencast is on YouTube~\cite{tool:YouTube}.
\end{abstract}



\maketitle

\renewcommand{\thefootnote}{*}
\footnotetext{These authors contributed equally to this work.}
\renewcommand{\thefootnote}{\arabic{footnote}}

\section{introduction}\label{sec:introduction}
Long methods encapsulate multiple responsibilities and are challenging to comprehend, debug, reuse, and evolve~\cite{Tsantalis2011, banker1993software, fowler1997refactoring}.
They often lead to code that is not only difficult to understand but also error-prone, thereby becoming a significant source of technical debt in software projects~\cite{fowler1997refactoring}. 
To alleviate this, developers frequently use the \textit{Extract Method} refactoring, which divides methods into smaller, more manageable units. This refactoring consistently ranks among the top five most commonly performed in practice~\cite{negara2013, murphy2012, refactoringminer2}. 

The \EM refactoring process comprises two phases: 
\begin{enumerate*}[label=(\roman*)]
\item Selecting statements for extraction from the original method, involving a meticulous examination of code segments that encapsulate specific responsibilities or logic, with the goal of isolating them into a distinct method,
\item Applying the refactoring, involving moving the selected statements into a brand new method, passing necessary variables as parameters, and invoking the new method from the original context.
\end{enumerate*}
While the application part has been a staple feature of all modern IDEs, they leave it up to developers to choose which  statements to extract.
The research community has developed diverse techniques to suggest statements for extraction.
Some research tools use static analysis in conjunction with software quality metrics~\cite{Maruyama2001, Tsantalis2011, Charalampidou2017, Yang2009APSEC, Tiwari2022ISEC}, such as statement cohesion~\cite{Tsantalis2011}, while others employ machine learning-based classifiers~\cite{gems2017ISSRE, REMS2023ICPC}


While the existing tools adhere to software quality metrics based on software engineering principles, they often generate suggestions that do not align with real-world \EM instances. 
We attribute this to the fact that refactoring requires both technical rigor and subjective discernment. 
In other words, developers rely on their understanding of software engineering principles and their subjective interpretation of code context when deciding what makes a good method. 
While existing techniques excel in the former, they often fall short in the latter. 
This gap may explain developers' reluctance to use automated refactoring support~\cite{WhyWeRefactorFSE2016}.

To bridge the gap and generate \EM suggestions that developers are likely to accept, we utilize Large Language Models (LLMs). 
They are trained on extensive code repositories containing millions of methods authored by actual developers. Thus, 
they are more likely to mimic human behavior and replicate how developers create methods, making them likely to propose refactorings that developers would embrace~\cite{dilhara2024unprecedented,OurTechReport2}. 
Our formative study, conducted on an extended corpus size of \extendedCorpusSize \EM real refactorings from open-source systems, demonstrated that LLMs are highly effective in providing expert suggestions, generating \averageSuggestionsPerFunctionCommunityCorpora suggestions per method.
This highlights the substantial potential of integrating LLMs into the domain of code refactoring. 
Nevertheless, LLM output cannot be directly adopted.
{Our findings indicate that \llmReliabilityPercentageExtendedCorpus of LLM suggestions are \emph{hallucinations} of two kinds. First, \llmInvalidSuggestionsPercentageExtendedCorpus of the LLM suggestions are invalid (e.g., syntactically incorrect), potentially causing compilation errors. Second, \llmNotUsefulSuggestionsPercentageExtendedCorpus of the suggestions are not useful (e.g., suggest to extract the entire method body).}


To advance the field of refactoring, we introduce \tool, the \emph{first automated refactoring tool to use LLMs}. \tool is an IntelliJ IDEA plugin. It employs LLMs to generate refactoring suggestions, then filters those that are invalid and non-useful, and further enhances valid suggestions using program slicing techniques. 
It then ranks them to offer high-quality options to developers.
Moreover, \tool bridges the gap between suggesting and applying refactorings by encapsulating selected suggestions into refactoring commands and executing them within the IDE. This process leverages the IDE to ensure the safe execution of refactorings, providing a seamless transition from suggestion to execution.

In~\cref{sec:effectiveness}, we first quantitatively evaluate the \tool in terms of its effectiveness compared to baseline tools. Then in~\cref{sec:user_eva} we qualitatively evaluate 
\tool's usability. 
In evaluations, we observed that EM-Assist really shines over previous approaches when replicating real-world refactorings in contemporary codebases.
When replicating \extendedCorpusSizeToRunCompetitors actual \EM refactorings that took place in open-source projects, EM-Assist’s recall rate was \extendedCorpusRecallEMAssist, compared to \extendedCorpusRecallJExtract for the previous best-in-class tool that relies solely on static analysis (JExtract).
For our qualitative evaluation (\cref{sec:user_eva}), we surveyed  \usabilityRespondents industrial developers, confirming the tool's usability at an approval rate of \surveyQoneApprovalRatio.


\section{\tool}


\begin{figure*}
\includegraphics[width=0.8\textwidth]{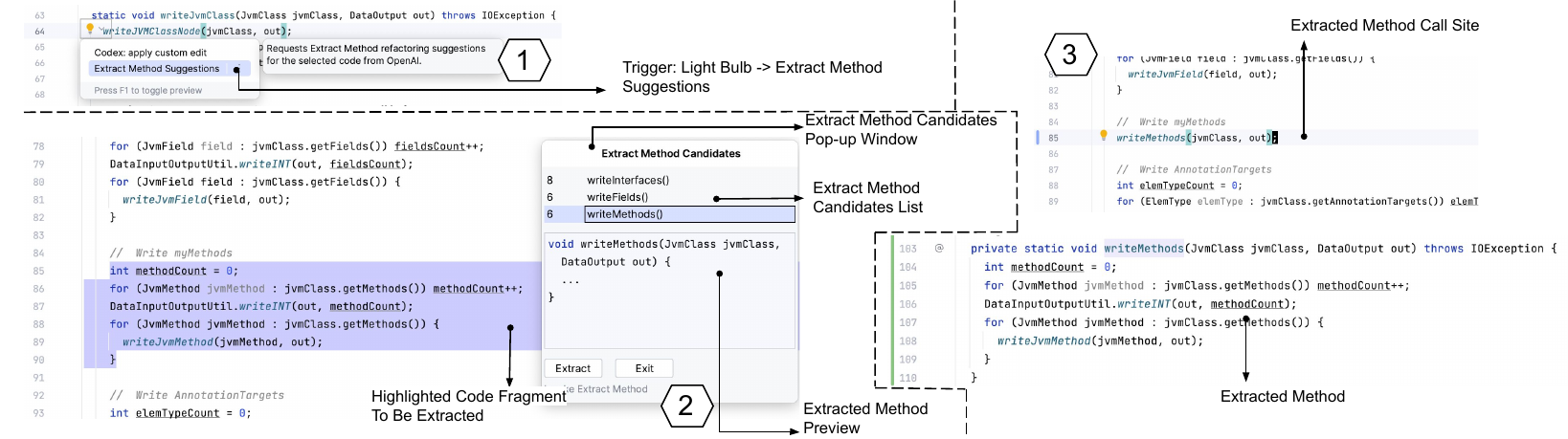}
    \caption{Workflow for using \tool within IntelliJ IDEA: 1) the user triggers the plugin to generate suggestions, 2) the plugin displays three refactoring options in a popup window, 3) the user selects one of the options and inspects the final code.}
    \label{fig:example2}
    \vspace{-.5cm}
\end{figure*}

\begin{figure}
\includegraphics[width=0.4\textwidth,keepaspectratio]{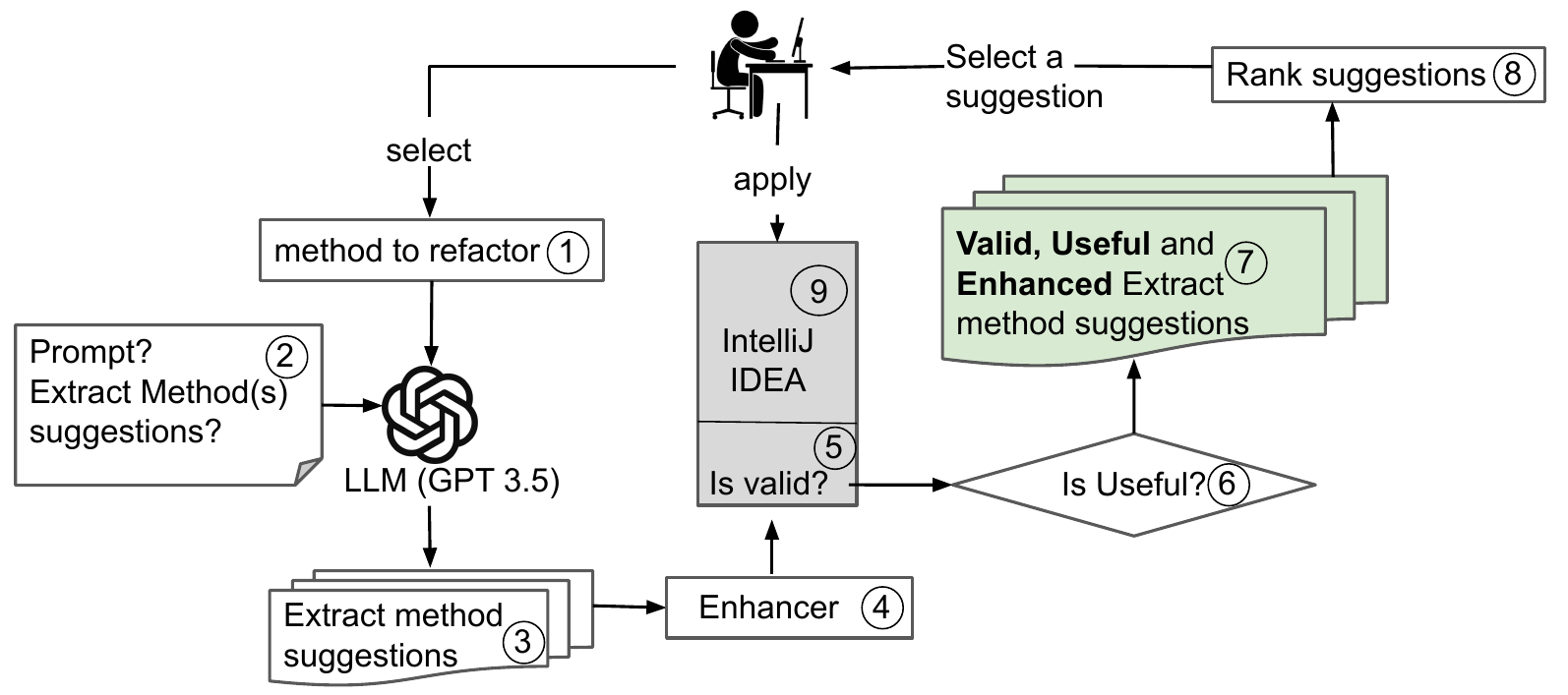}
    \caption{The workflow of generating refactoring suggestions}
    \label{fig:workflow}
    \vspace{-.5cm}
\end{figure}

\subsection{Example Usage of \tool}
\Cref{fig:example2} shows an example of using \tool to suggest and execute refactorings on the method \smcode{writeJvmClass()} within the project \textit{JetBrains/intellij-community}. 
The user triggers \tool by choosing the ``Extract method suggestions'' option from the intention actions (yellow light bulb)\footnote{https://www.jetbrains.com/help/idea/intention-actions.html} for the  method (\circled{1} in \Cref{fig:example2}).
In a popup window, \tool shows three distinct suggestions for extracting code fragments, accompanied by a suggested method name (\circled{2} in \Cref{fig:example2}). 
In this example, the plugin suggests three extract methods: \smcode{writeInterfaces()}, \smcode{writeFields()}, and \smcode{writeMethods()}, and it indicates the size of the code fragment for each suggestion. 
To make it convenient for the user to assess each suggestion, \tool helps the user preview the signature of the proposed new method (including method parameters) directly within the popup window.
Additionally, for each suggestion, \tool highlights the  code fragment slated for extraction. 
For example, the blue-colored statements in \Cref{fig:example2} (line 85-90 of \circled{1}) represent the proposed statements for extraction within the \smcode{writeMethods()} suggestion. 
Likewise, the plugin affords flexibility to users, allowing them to choose the most appropriate extract method suggestion and seamlessly apply it to the \smcode{writeJvmClass()} method. 
To do so, the user clicks the ``Extract'' button conveniently located within the popup window.
In this example, the user chose to apply the \smcode{writeMethods} suggestion to the code, and the resulting code is shown as \circled{3} in \Cref{fig:example2}.

\subsection{Workflow}
In this section, we present the internal workflow that \tool uses to
automatically suggest and perform \EM.
As shown in \Cref{fig:workflow}, the developer invokes the plugin by clicking the ``Extract Method'' option in the intention actions for a particular method they intend to apply the extract method refactoring to 
(\circled{1} in \Cref{fig:workflow}).
Then \tool creates a LLM few-shot learning prompt (\circled{2} in \Cref{fig:workflow}) that contains the method declaration, and it invokes the LLM programmatically to generate refactoring suggestions (\circled{3} in \Cref{fig:workflow}). To tame the non-determinism of LLMs, \tool repeats the same prompt request, using optimal values that we determined empirically (i.e., between 5-10 iterations). 
Then \tool enhances the LLM suggestions by adjusting the code fragments, using program slicing (\circled{4} in \Cref{fig:workflow}).
However, not all these enhanced suggestions are safe to apply, as some may result in non-compilable and erroneous code.
In fact, our analysis conducted on a dataset of \extendedCorpusSize methods, reveals that  \llmReliabilityPercentageExtendedCorpus of the suggestions generated by the LLM are invalid, potentially resulting in non-compilable code.
To filter invalid suggestions, \tool employs the IntelliJ IDEA APIs for determining whether a code fragment meets the refactoring preconditions for extract method (\circled{5} in \Cref{fig:workflow}).

Some valid suggestions are not practical for real-world implementation (e.g., those that contain the whole method body). 
Therefore, \tool filters unuseful suggestions (\circled{6} in \Cref{fig:workflow}).
In the end, \tool generates numerous suggestions per long method (\circled{7} in \Cref{fig:workflow}), which could potentially overwhelm the user if all are presented at once. 
To address this, we employ a ranking mechanism (\circled{8} in \Cref{fig:workflow}) that relies on the frequency of LLM suggestions.
The ranking mechanism is centered on the idea that if a particular code fragment is repeatedly suggested for extraction by the LLM during subsequent triggers, it must represent an important functional concern that is suitable to be in a separate method.
Then, \tool presents the ranked suggestions (top-n, default is top-3) to developers through our user interface, offering previews of extracted method signatures and associated code fragments. 
After the user selects their preferred suggestion, \tool encapsulates it in a refactoring command, 
seamlessly invoking IntelliJ IDEA to execute the refactoring correctly (\circled{9} in \Cref{fig:workflow}).


\subsection{Implementation}
\tool is implemented as an IntelliJ IDEA plugin and works as an intention action. IntelliJ Platform has a suite of APIs for Abstract Syntax Tree (AST) manipulation and refactoring. \tool leverages IntelliJ's AST manipulation through the Program Structure Interface (PSI), which is the layer in the IntelliJ Platform responsible for code, file, and project model. \EM refactoring is applied through IntelliJ's refactoring framework. We employ few-shot learning~\cite{gpt-3,radford2019language} to form the LLM prompt.  \tool is configurable to use various LLMs, by default we use OpenAI's GPT-3.5-turbo. The communication with LLM is asynchronous, thus, the IntelliJ code editor is not frozen and multiple requests are sent to the LLM in parallel. Not all suggestions provided by the LLM are valid and ready to be extracted, therefore, we use PSI elements to adjust the scope of these \EM candidates. We utilize IntelliJ's Extract Method framework to further filter out the candidates which cannot be extracted. We enhanced the Extract Method framework to support custom method names suggested by the LLM.

\textbf{Extensions:} \tool currently supports Java and Kotlin, but its implementation offers an extensible framework for supporting other languages as well. For example, to support \EM in Python code, extensions would be needed in steps \circled{1}, \circled{3}, \circled{5}, and \circled{7} (\Cref{fig:workflow}), whereas \circled{2}, \circled{4}, \circled{6}, \circled{8}, \circled{9} (\Cref{fig:workflow}), could be reused. To support other IDEs that provide good support for ExtractMethod refactoring, steps \circled{5} and \circled{9} (\Cref{fig:workflow}) would need change, whereas all remaining steps could be reused. Moreover, our preliminary experiments with using LLMs to suggest other kinds of refactorings show that LLMs are prolific, but we need to employ a  similar approach: use the LLM for developer-aligned suggestions, but carry out the refactoring plan with the safety of the IDE.

{\textbf{Limitations:} Currently, using \tool requires sending the user's host method code to OpenAI. As a result, the response time can vary. Our experiments involving 25,235 invocations of the tool revealed that, on average, \tool takes 2 seconds to suggest refactorings. This  includes the time taken to invoke the LLM and the additional processing time required by the tool. According to Nielsen's usability guidelines~\cite{nielsen1994usability}, this response time is acceptable for human-operated software tools, ensuring that the user's flow of thought remains uninterrupted.}
Moreover, when  developers work on proprietary code, or a company policy prohibits using Generative AI solutions, this currently limits the use of \tool. Emerging solutions such as deploying LLM servers fully on-premise (for enterprise customers or government organizations) or using medium-size models locally, could make Generative AI-based solutions even more broadly applicable.
{Moreover, \tool's architecture make it easy to switch between different LLMs.}

\section{evaluation}
We conducted a comprehensive evaluation of \tool, using a harness of two complementary approaches.
\begin{enumerate*}[label=(\roman*)]
  \item {Quantitative (\cref{sec:effectiveness}): we compared the effectiveness of EM-Assist with two state-of-the-art extract method recommendation tools by replicating a large dataset of \extendedCorpusSizeToRunCompetitors actual extract methods performed by open-source developers}.
  Unlike previous studies that used a small, synthetic corpus of artificially-created ExtractMethods, we are replicating real-world refactorings from famous, contemporary open-source projects. 
  We believe this is more indicative of a tool's capabilities in dealing with the complexities of real-world refactorings, and the large scale of experiments guards against over-fitting a tool for a small corpus. 
  \item Qualitative (\cref{sec:user_eva}): we evaluated the usability of \tool's approach when guiding industrial developers to follow our novel workflow for performing ExtractMethod refactorings. Unlike previous studies that have employed large number of students, we believe our study employing 18 full-time, experienced professional developers is more indicative of the quality of the suggested refactorings.   
\end{enumerate*}

\subsection{Effectiveness of \tool}\label{sec:effectiveness}

{To determine \tool's effectiveness in recommending refactorings that align with developer preferences, we compare it to previous state-of-the-art approaches that rely solely on static analysis and software quality metrics. 
In our technical report~\cite{OurTechReport2}, we compare \tool to six other tools that suggest \EM and we use a synthetic corpus used by other researchers. We found that \textsc{JExtract}~\cite{jextractsilva2015} outperforms the previous tools, thus we use \textsc{JExtract} as the strongest competitor to \tool.}



First, we construct an oracle. We mined the complete version history of  projects using \textsc{RefactoringMiner}~\cite{refactoringminer2}, a state-of-the-art tool for refactoring detection.
Thus we identified \extendedCorpusSizeToRunCompetitors cases when the open-source developers performed ExtractMethod refactorings. 
We then replicate all these refactorings by running the tools on the original host methods (the version before refactoring was applied) and compare the suggested refactorings against the actually-performed refactorings in the oracle.
The oracle exhibited a diverse range of method involved in refactoring, with min/max/mean/median values of 3/1494/30/5 LOC for the host method, and  2/95/6/3 LOC for extract methods. 
Given the size of host methods, there can be many ways to group code fragments for extraction.

{We executed the two tools on each host method in the oracle. We then compared the refactorings performed by these tools and cross-referenced them with the oracle to ensure that the extracted code lines matched those in the oracle.
Following best practices established in prior research~\cite{jextractsilva2015,Tsantalis2011,Charalampidou2017,fernandes2022live,REMS2023ICPC,gems2017ISSRE} 
we evaluate top-5 suggestions generated by the tools, calculating Recall@5 at tolerance level 3\%.
{We selected Recall as the performance metric as this is the metric used by 
all previous tools.}
Given the size of host methods, even at 3\% tolerance, it means in many cases we must have an exact match between one of the top-5 suggestions and the extracted code fragment in the oracle. 
We compute the recall by dividing the total number of refactoring suggestions that matched the oracle to the total number of extract methods in the oracle. 

\tool has a recall rate of {\extendedCorpusRecallEMAssist}, significantly surpassing its counterparts whose best recall was \extendedCorpusRecallJExtract. 
Acknowledging the probabilistic nature of \tool's underlying LLM, we conducted further statistical analysis for robust validation. 
We repeated each prediction from \tool 50 times and performed a one-sample t-test against \textit{JExtract}, the second-best performing tool. The test decisively rejected the null hypothesis asserting that \textit{JExtract} had a better recall rate than EM-Assist, with a p-value of $<10^{-5}$.}


\subsection{Usability of \tool}\label{sec:user_eva}
We complement the previous quantitative evaluation with  a survey to evaluate developers' perception of using the novel workflow of \tool in comparison to the current workflow in the IDEs.
We intentionally designed the survey to allow participants to view a brief, feature-focused demo of the plugin, rather than requiring them to install and execute it. Previous research~\cite{davidoff2007rapidly, kery2020mage} demonstrated that this approach is useful for evaluating the fundamental concepts and ideas behind a tool, as high-fidelity prototypes might overwhelm survey participants with excessive implementation details.

We surveyed \usabilityRespondents qualified professional software developers.
Among these participants, 50\% have more than a decade of experience in software development, 33\% have between one and three years of experience, while others have between three and seven years of experience.
We used Likert-type questions~\cite{boone2012analyzing} to evaluate \tool's features, 
and we included open-ended questions to capture their insights regarding future enhancements. 
We adhered to established best practices in usability studies~\cite{barnum2010usability}.

\begin{figure}[h]
\includegraphics[width=0.44\textwidth,keepaspectratio]{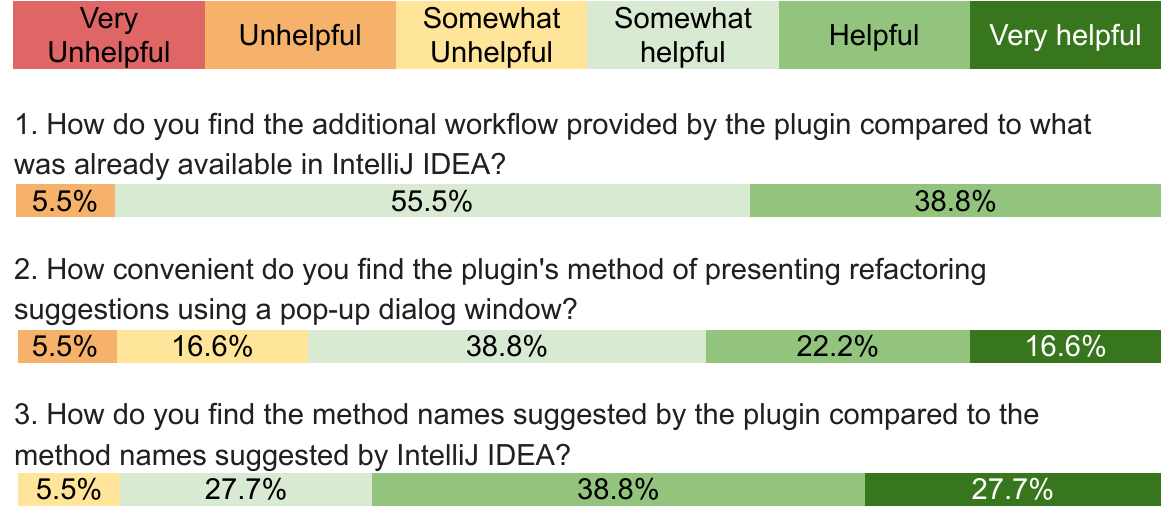}
    \vspace{-0.35cm}
    \caption{Results of the conducted usability survey.\label{fig:usability}} 
    \vspace{-0.5cm}

\end{figure}

~\Cref{fig:usability} shows the results of the Likert-type questions. 
The respondents expressed highly positive attitudes towards the overall usability and user interfaces of \tool.
\qOneAcceptance of the respondents found the additional workflow stages, including triggering the plugin, generating LLM suggestions, and selecting one to apply, to be convenient and easy to handle.
\tool uses a popup window to present the suggestions, and over \qTwoAcceptance of respondents agreed that it is somewhat helpful or greater.
Additionally, \qThreeAcceptance of respondents liked the suggested method names. While users could get name suggestions from an LLM at a later time, we conveniently offer meaningful names at the time of performing extract method.

In the open-ended questions, the respondents suggested valuable enhancements to \tool in terms of usability, configurability, and mode of execution, which we will use in a future release.

Overall, we received positive praises from the survey respondents about \tool:
\textit{``It looks super cool so far! :fire:''}. Another said: 
\textit{``Thank you for interesting suggestions! Hope to see this in production in the future.''} And another: 
\textit{``...these suggestions made me look at this code with
new eyes once more, and I will try to refactor it.''}

\section{related work}
Researchers developed tools to suggest code fragments to extract.
Many primarily rely on static analysis-based rules like program slicing~\cite{lakhotia1998restructuring, Maruyama2001, abadi2009fine}, the separation of concerns~\cite{silva2014recommending}, or the single responsibility principle \cite{Charalampidou2017}. While these tools excel at adhering to software quality principles, they face a significant limitation.  They are not able to adapt suggestions to developers' subjective intentions, possibly because of their lack of access to real-world data illustrating actual developer refactoring practices. 

As an alternative to rule-based techniques, machine learning-based classifiers~\cite{REMS2023ICPC, gems2017ISSRE} have been proposed for refactoring suggestions.  
Despite having access to relevant data, they grapple with practical challenges such as data scarcity, protracted training periods, and resource-intensive training procedures. 
Furthermore, these ML models are often specific to particular tasks, necessitating periodic retraining to maintain their effectiveness~\cite{software2,pyevolve,mlrepetitive2,mlrepetitive1}. 
In contrast to these existing tools, \tool leverages LLMs as a promising solution. 
LLMs are general-purpose models pre-trained on extensive source code and text, enabling them to generate refactoring suggestions aligned with developer intentions using knowledge from both code and textual documents. 
\tool takes this a step further by removing hallucinations, validating and enhancing the LLM output, and providing developers with tailored recommendations..

While obtaining the training set for every kind of refactoring is a formidable challenge for previous approaches, it is not for LLM-based tools. While previous approaches require continuous/custom model training or custom analysis, given the broad knowledge-base of LLMs, 
our solution is easier to use, maintain, and deploy in IDEs. 

\section{Conclusions}
We present \tool, the first automated refactoring tool that employs LLMs, implemented as 
an IntelliJ IDEA plugin. It bridges the gap between suggesting and applying \EM refactoring, and shrinks the gap between refactoring suggestions and developer practices. 
\tool generates refactoring suggestions using LLMs, then validates and enhances them with static analysis in the IDE. To avoid overwhelming developers, \tool ranks and presents only the highest-quality suggestions, and then correctly executes the user-selected refactoring suggestion with the IDE.
We comprehensively evaluate \tool and it outperforms state-of-the-art. Additionally, a developer survey 
confirmed the high usability of the plugin.
\tool uses a novel way to check LLM results and make them useful for refactoring tasks. Being the first tool to employ LLMs for \EM refactoring, \tool demonstrates the potential of LLMs as effective refactoring assistants. 
{We are now expanding \tool to support many more kinds of refactorings.}

\section{Acknowledgements}
{We thank the ML Methods in Software Engineering Lab at JetBrains Research, and the FSE-2024 reviewers for their insightful and constructive feedback for improving the paper.
This research was partially funded through the NSF grants CNS-1941898, CNS-2213763, and the Industry-University Cooperative Research Center on Pervasive Personalized Intelligence.}

\balance
\bibliographystyle{ACM-Reference-Format}
\bibliography{references_with_all_names.bib}

\end{document}